\documentclass[
reprint,showpacs,preprintnumbers, amsmath,amssymb,aps,prl]{revtex4-1}
\usepackage{graphicx}
\usepackage{dcolumn}
\usepackage{bm}
 \usepackage{natbib}
\usepackage{subfig}
\usepackage{placeins}
 \usepackage[pdftitle={Optical Absorption in B$_{13}$ Cluster: A Time-Dependent Density Functional Approach},
pdfauthor={Ravindra Shinde, Meenakshi Tayade}]{hyperref}
  \hypersetup{
  unicode=false,          
  pdftoolbar=true,        
  pdfmenubar=true,        
  pdffitwindow=true,     
  pdfstartview={FitH},    
  pdfsubject={Computational Condensed Matter},   
  pdfnewwindow=true,      
  pdfcreator={RevTeX},
  colorlinks=true,       
  linkcolor=red,          
  citecolor=blue,        
  urlcolor=blue,           
  }
\usepackage[all]{hypcap}
\begin{document}

\preprint{AIP Conference Proceedings (G-163)}

\title{Optical Absorption in B$_{13}$ Cluster: A Time-Dependent Density Functional Approach}

\author{Ravindra \surname{Shinde}}
\email[Corresponding author: ]{ravindra.shinde@iitb.ac.in}
\affiliation{Department of Physics, Indian Institute of Technology Bombay, Mumbai 400076, Maharashtra, INDIA.}

\author{Meenakshi \surname{Tayade}}
\email{mkstayade@gmail.com}
 \affiliation{Department of Chemistry, Institute of Chemical Technology, Mumbai 400019, Maharashtra, INDIA. }%
\date{October 11, 2012}

\begin{abstract}
The linear optical absorption spectra of three isomers of planar boron cluster B$_{13}$ are calculated using 
time-dependent spin-polarized density functional approach. The geometries of these cluster are optimized at the 
B3LYP/6-311+G* level of theory. Even though the isomers are almost degenerate, the calculated spectra are quite different,
indicating a strong structure-property relationship. Therefore, these computed spectra can be used in the 
photo-absorption experiments to distinguish between different isomers of a cluster.
\end{abstract}


\maketitle

\section{Introduction}

Boron clusters exhibit novel properties and have wide range of applications, from hydrogen storage to 
chemical ligands.\cite{1} The ability to form structures of any size makes boron different from other elements.

Recent reports have shown that the planar boron clusters are analogous to hydrocarbons with both $\sigma$ and
$\pi$ aromaticity.\cite{2,3} For example, wheel shaped B$_{19}^{-}$, shows chemical bondings similar to
[10]annulene(C$_{10}H_{10}$) and coronene(C$_{24}H_{12}$).\cite{3} ab initio study of another planar cluster,
 B$_{13}^{+}$, reveals that this cluster act as a Wankel motor, i.e. the outer ring rotates in opposite 
direction as that of the inner, when shined with circularly polarized light. \cite{4}

In this report, we present a theoretical calculation of optical absorption spectra of the most stable planar
isomers of B$_{13}$ cluster. Since, conventional mass spectrometry alone cannot distinguish between the
isomers of a given cluster; our results will be useful in identifying the different isomers, with the help of the
computed absorption spectra. We have used this method earlier to study the optical absorption in other
smaller boron clusters.\cite{1}

\section{Computational Details}
\begin{figure}[h!]
 \centering
 \includegraphics[width=8cm]{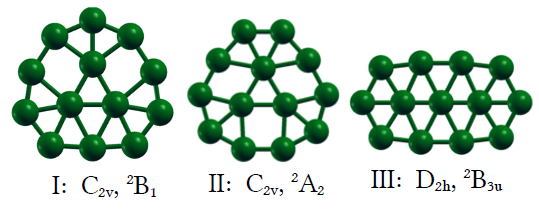}
 \caption{The optimized geometry, point group symmetry and electronic state of the three isomers of B$_{13}$ cluster.}
 \label{fig:combined_b13}
\end{figure}

The geometries of different isomers were optimized at the B3LYP/6-311+G* level of theory. \cite{5}
These geometries were used to calculate the optical absorption spectra. Fig. \ref{fig:combined_b13} shows the optimized
structures, the point group symmetries and ground state electronic state.

The absorption spectra were calculated using the time-dependent spin-polarized density functional
theory (TDDFT). Calculations are performed at zero temperature and fixed geometries. Norm-conserving
pseudopotentials and the PBE parameterization are employed in the adiabatic approximation for the
exchange-correlation potential. 
\newpage
\section{Results}

\begin{figure}[h!]
 \centering
 \subfloat{\includegraphics[width=8cm]{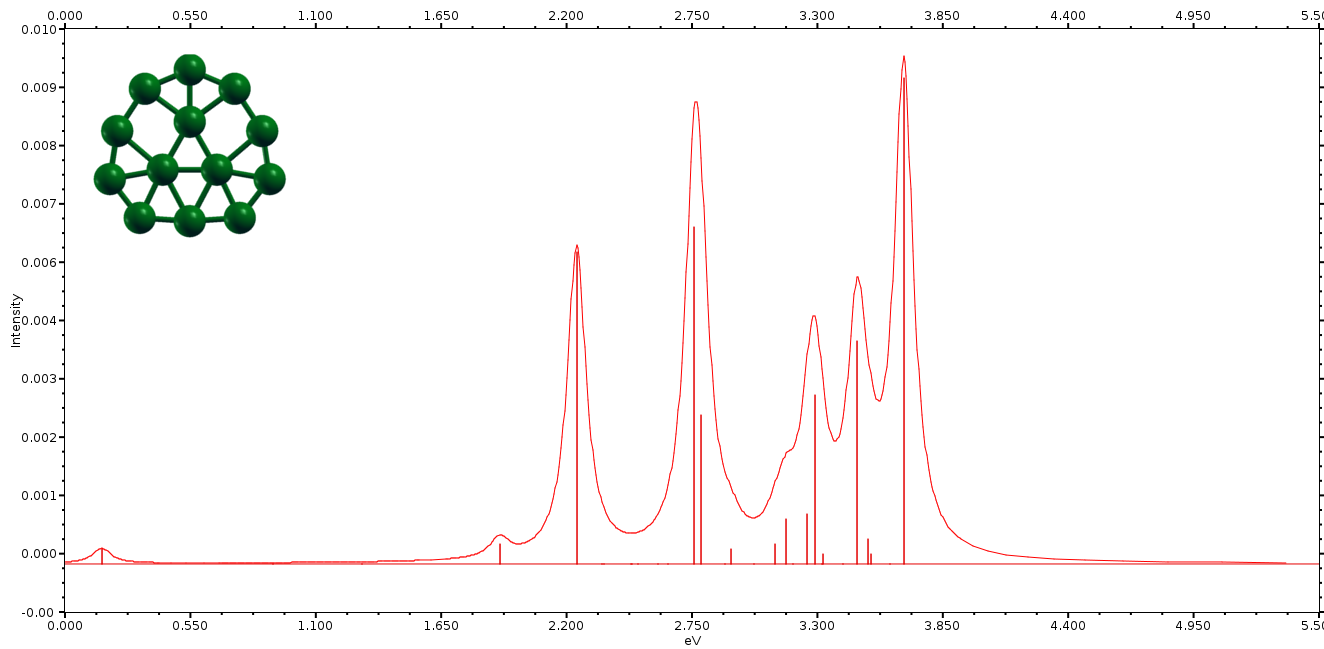}} \\
 \subfloat{\includegraphics[width=8cm]{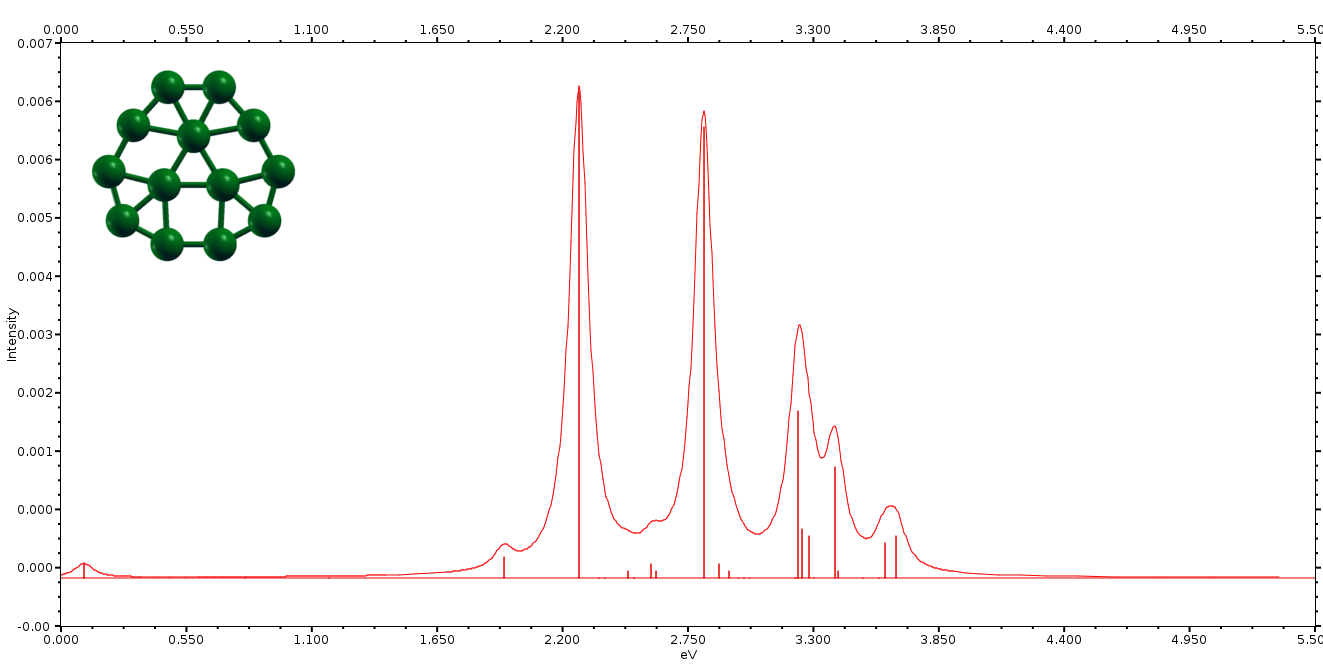}} \\
 \subfloat{\includegraphics[width=8cm]{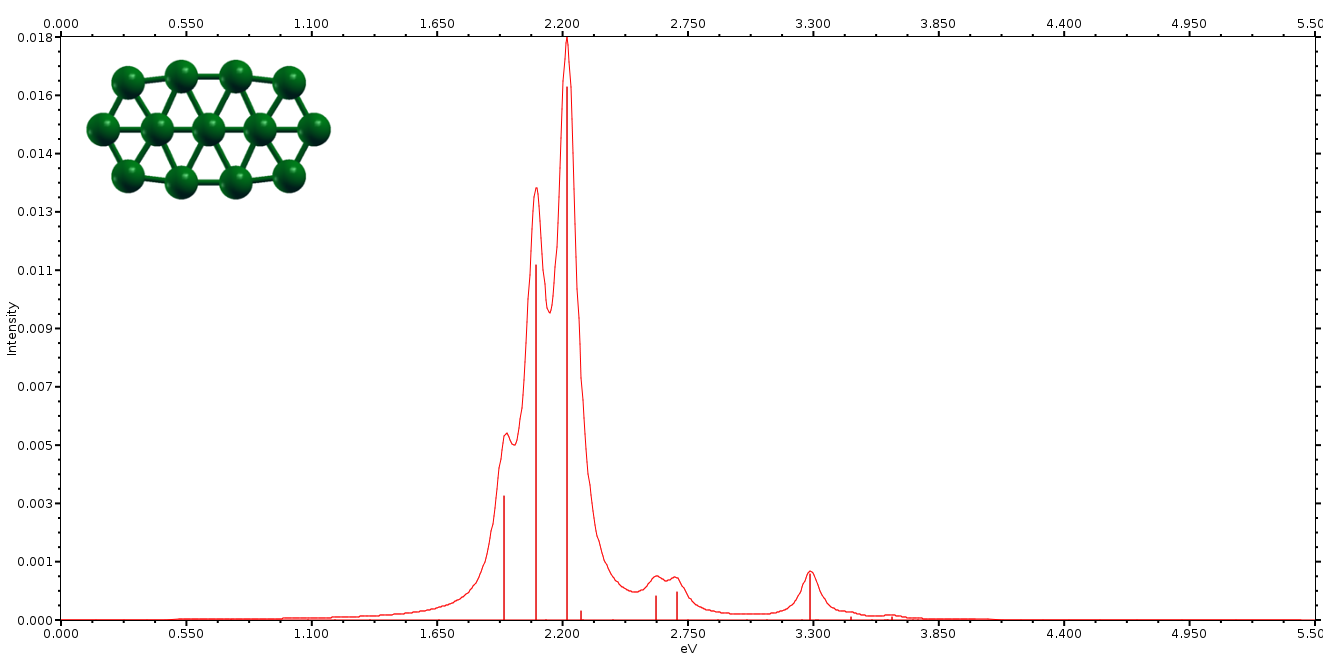}}
\captionsetup{format=default,justification=centerlast}
\caption{The calculated TD DFT optical absorption spectra of isomers of B$_{13}$ cluster. An artificial line width of
0.1 eV is used throughout. Intensities are in arbitrary units. The vertical lines denote the calculated spectra, 
without the Gaussian convolution.}
\label{fig:plots}
\vspace{-2em}
\end{figure}

The relative ground state energies of three isomers of B$_{13}$  cluster are 0.0 eV, 0.015 eV and 0.30 eV
respectively. This difference is quite small, and is susceptible to the level of theory used in the
calculations. However, the optical absorption spectra of the isomers -- as shown in the Fig. \ref{fig:plots} -- are
completely different from each other.

The excitations involved in first four peaks of first two isomers are mainly characterized by $\sigma \rightarrow \pi^{*}$ 
 type of excitation. However, the isomer I shows more absorption in the higher energy range characterized by
$\pi \rightarrow \pi^{*}$, as compared to the second isomer, which shows excitation $\sigma \rightarrow \pi^{*}$
 and $\pi \rightarrow \sigma^{*}$. The spectrum of third isomer is quite different than the rest of the two,
as expected. It has a large optical gap and the bulk of the optical absorption is concentrated within a narrow
energy window. The most intense peak, at 2.2 eV, is characterized by $\pi \rightarrow \pi^{*}$ transition.

 \FloatBarrier

\begin{table}
\captionsetup{justification=centerlast}
 \caption{Excitation involved in major peaks of the optical absorption spectra of isomer I.} \label{tab:isomer1}
\begin{ruledtabular}
\begin{tabular}{ccl}
Sr. No. & Energy (eV) & Transition \tabularnewline
\hline
1 & 2.24 & $H-1 \rightarrow L$ \tabularnewline
  &      & $H-2 \rightarrow L$  \tabularnewline
  & 	 & $H-1 \rightarrow H$ \tabularnewline
\tabularnewline
2 & 2.78 & $H-2 \rightarrow L$  \tabularnewline
  &  	 & $H   \rightarrow L+4$  \tabularnewline
\tabularnewline
3 & 3.68 & $H-4 \rightarrow L+1$  \tabularnewline
\end{tabular}
\end{ruledtabular}
\end{table}
\vspace{-2em}

\begin{table}
\captionsetup{justification=centerlast}
 \caption{Excitation involved in major peaks of the optical absorption spectra of isomer II.} \label{tab:isomer2}
\begin{ruledtabular}
\begin{tabular}{ccl}
Sr. No. & Energy (eV) & Transition \tabularnewline
\hline
1 & 2.27 & $H-1 \rightarrow L$ \tabularnewline
  & 	 & $H-1 \rightarrow H$ \tabularnewline
\tabularnewline
2 & 2.82 & $H-1 \rightarrow L+1$  \tabularnewline
  &  	 & $H-2   \rightarrow L$  \tabularnewline
\tabularnewline
3 & 3.23 & $H-4 \rightarrow L$  \tabularnewline
\end{tabular}
\end{ruledtabular}
\end{table}
 \vspace{-2em}

\begin{table}
\captionsetup{justification=centerlast}
 \caption{Excitation involved in major peaks of the optical absorption spectra of isomer III.} \label{tab:isomer3}
\begin{ruledtabular}
\begin{tabular}{ccl}
Sr. No. & Energy (eV) & Transition \tabularnewline
\hline
1 & 1.95 & $H-3 \rightarrow H$ \tabularnewline
  & 	 & $H-2 \rightarrow L$ \tabularnewline
\tabularnewline
2 & 2.08 & $H-2 \rightarrow L$  \tabularnewline
\tabularnewline
3 & 2.22 & $H-1 \rightarrow L$  \tabularnewline
\end{tabular}
\end{ruledtabular}
\end{table}
\vspace{-2em}
\section{Conclusion}
The optical absorption spectra of three planar isomers of the B$_{13}$ Cluster are calculated using TDDFT
approach. The nature of the optical excitation is investigated. This study may help in distinguishing the
experimentally produced isomers of B$_{13}$, which is not possible with conventional mass spectrometry alone.

\section{ACKNOWLEDGMENTS}
The author (R.S.) would like to acknowledge the CSIR, India, for their financial support. We thank A.
Shukla and A. Sergeeva for useful discussions.

\end{document}